\documentclass[preprint,showpacs,preprintnumbers,amsmath,amssymb]{revtex4}
\usepackage{graphicx}
\usepackage{dcolumn}
\usepackage{bm}

\begin{document}

\title{Anharmonic stacking in supercoiled DNA}

\author{Marco Zoli}

\address{
School of Science and Technology - CNISM \\  University of Camerino, I-62032 Camerino, Italy \\ marco.zoli@unicam.it}

\date{\today}

\begin{abstract}
Multistep denaturation in a short circular DNA molecule is analyzed by a mesoscopic Hamiltonian model which accounts for the helicoidal geometry. Computation of melting profiles by the path integral method suggests that stacking anharmonicity stabilizes the double helix against thermal disruption of the hydrogen bonds. Twisting is essential in the model to capture the importance of nonlinear effects on the thermodynamical properties. In a ladder model with zero twist, anharmonic stacking scarcely affects the thermodynamics. Moderately untwisted helices, with respect to the equilibrium conformation, show an energetic advantage against the overtwisted ones. Accordingly moderately untwisted helices better sustain local fluctuational openings and make more unlikely the thermally driven complete strand separation.
\end{abstract}



\maketitle

\section*{1. Introduction}

In the transcription process, the RNA polymerase attaches to the DNA near the start of the gene and opens a small segment of the double helix thus reading and coping the coded instructions into a template, the mRNA, for the synthesis of proteins. Both the molecular machine and the transcription bubble move along the gene by activation processes associated to proteins which bind to DNA portions, \emph{enhancers} \cite{black}, located even several thousands bases away along the chain with respect to the \emph{promoter} starting region \cite{nolis}. In some cases, long range enhancer-promoter interactions appear to be mediated by DNA structural deformations such as looping, bending and twisting whose molecular origin is under investigation. Certainly the flexibility of the double helix favors long range effects in biological functioning \cite{englan,yera,weres} whereas twist deformations modify the three dimensional structure, producing compact supercoiled shapes which are crucial for DNA packaging within the nucleus  \cite{gore,marko2,gaeta,shei}.
It is known that the opening of transcription bubbles is possible only if accompanied by a local decrease in the twist angle with respect to the equilibrium value. Unwinding is measured by a \emph{negative supercoiling } and the undertwisting of DNA can yield concentration of energy such to break hydrogen bonds, eventually leading to denaturation  \cite{alle,metz10}.
Twisting is described by the angle that the base pairs rotate around the molecule axis. B-DNA at room temperature has a helix repeat of $\sim 35 \,{\AA}$ hosting $\sim 10.4$ base pairs, hence the equilibrium twist angle is $\theta_{eq} \sim 0.6 \,rad$.
As proposed by Prohofsky \cite{proh}, such helical structure may be the optimal configuration to capture the chemical energy flux which nourishes RNA polymerase in DNA transcription. Similar arguments have been recently raised by a numerical study of the nonlinear dynamics in helicoidal DNA showing that the energy localization required for bubble formation is driven by anharmonic stiffness \cite{tabi}. In this regard, a microscopical understanding of $\theta_{eq}$ in unperturbed DNA seems therefore preliminary to the comprehension of those local fluctuational openings occurring in replication and transcription events \cite{benham,segal}. As the latter involve disruption of the base pairs hydrogen bonds in a way similar to the denaturation process, whether mechanically or thermally driven \cite{siggia,cocco,hanke,singh}, the properties of the melting transition have drawn such a great interest over the last decades \cite{poland,fisher,azbel1,wart,cule,peliti,stella,yakus,joy05,manghi,jost,santos1}.

To elucidate the DNA biological functions in terms of the fundamental interactions, path integral techniques \cite{io,io1} are applied in this paper to the thermodynamics of a circular molecule described by a mesoscopic nonlinear Hamiltonian, the Dauxois-Peyrard-Bishop (DPB) model \cite{pey1,pey2}.
Hamiltonian approaches have the advantage of treating the system at the level of the base pair \cite{salerno} thus permitting to study the thermodynamic behavior by tuning the heterogeneity degree in the sequence.

In a previous investigation \cite{io11}, the standard DPB model has been enriched by introducing a rotational term in the stacking potential thus accounting for the twist between adjacent base pairs along the molecule axis. It has been found that the equilibrium twist $\theta_{eq}$ provides in fact the stablest conformation against thermal disruption of the base pair bonds in the denaturation regime. As the twist on its own gives stability to the system, it has consistently been shown that the melting profiles of the twisted conformation are less affected by changes in the harmonic backbone coupling than the untwisted DPB model. However no analysis of the anharmonic stacking effects has been carried out in Ref.\cite{io11}. As the DPB model attributes a special role to the anharmonic backbone interactions in sharpening the denaturation transition \cite{theo}, here I study specifically the relation between anharmonic stacking and twist angle in a modified DPB model which also includes a stabilizing solvent interaction term. It is found that the twisting of the double helix is essential to switch on the anharmonic stacking effects on the melting profiles whereas stacking anharmonicity is negligible in the standard DPB model.

The system Hamiltonian is presented in Section 2 together with a discussion of the stacking potential energetics as a function of the twist. Section 3 outlines the path integral computational method. Section 4 contains the results: the melting profiles are displayed both for underwound and overwound short double helices while the effects of stacking anharmonicity are reported for the equilibrium twist case and for the untwisted DPB model. Also the role of the solvent potential is discussed. Some final remarks are made in Section 5.

\section*{2. Model}

To model a double helix of $N$ base pairs with reduced mass $\mu$, I start from a generalized DPB Hamiltonian in which the pair mates separation, with respect to
the ground state position, is given by the transverse stretching $y_n$ (for the \emph{n-th} base pair). These are generally much larger than the longitudinal base pairs displacements along the molecule backbone which are accordingly dropped. Thus, the one-dimensional Hamiltonian reads:

\begin{eqnarray}
& & H =\, \sum_{n=1}^N \biggl[ {{\mu \dot{y}_{n}^2} \over {2}} + V_M(y_n) + V_{sol}(y_n) +  V_S(y_n, y_{n-1}) \biggr] \, \nonumber
\\
& & V_M(y_n) =\, D_n \bigl(\exp(-a_n y_n) - 1 \bigr)^2  \, \nonumber
\\
& & V_S(y_n, y_{n-1})=\, {K \over 2} g_{n,n-1}( y_n^2 - 2 y_n y_{n-1}\cos\theta + y_{n-1}^2 ) \, \nonumber
\\
& & g_{n,n-1}=\,1 + \rho \exp\bigl[-\alpha(y_n + y_{n-1})\bigr]\, \, \nonumber
\\
& & V_{sol}(y_n) =\, - D_n f_s \bigl(\tanh(y_n/\l_s) - 1 \bigr)  \,
.
\label{eq:1}
\end{eqnarray}

Like the DPB Hamiltonian,  Eq.~(\ref{eq:1}) incorporates nonlinearities \emph{both} in the hydrogen bond base pair interactions shaped by the Morse potential $V_M(y_n)$ \emph{and} in the stacking coupling $V_S(y_n, y_{n-1})$ between neighboring bases along the two strands.
$D_n$ and $a_n$ are the pair dissociation energy and the inverse length setting the hydrogen bond potential range: they account for heterogeneity in the sequence.

Considering that: \emph{i)} the AT- and GC-base pairs are linked by two and three hydrogen bonds respectively, \emph{ii)} there is electrostatic repulsion between the highly charged phosphate groups on complementary strands, the pair dissociation energies can be reasonably taken as
$D_{AT}=\,30 meV$ and $D_{GC}=\,45 meV$. These values are above $k_BT$ at room temperature.  $k_B$ is the Boltzmann constant and $T$ is the temperature. Larger energy values \cite{campa,theo1} would shift the denaturation steps at higher $T$. The inverse lengths are $a_{AT}=\,4.2 {\AA}^{-1}$ and $a_{GC}=\,5 {\AA}^{-1}$ in view of the fact that AT-base pairs have larger displacements than GC-base pairs.

$K$ is the harmonic stacking related to $\mu$ by $K=\,\mu \nu^2$, $\nu$ being the frequency of the phonon mode. $\rho$ and $\alpha$ are the anharmonic stacking parameters which are also assumed independent of the type of base at the $n$ and $n-1$ sites.
The homogeneity assumption \cite{ares1} for the stacking potential relies on the observation that both types of base pairs contain a purine plus a pyrimidine, the former being larger and heavier. Thus the AT- and GC- base pairs are comparable in size and weight with their effective masses usually taken as $\mu \sim 300 a.m.u.$.
While the inclusion of stacking heterogeneity is motivated in a growing number of studies \cite{bonnet,yakov,metz06,cooper} and may be straightforwardly incorporated in the path integral approach, there is no need to further enlarge the set of input parameters to the purposes of the present work.

I take values in the commonly used range, $K=\, 60meV {\AA}^{-2}$,  $\rho=\,2$,  $\alpha=\,0.5 {\AA}^{-1}$ whereas the anharmonicity effects due to larger $\rho$ are discussed below.  Somewhat smaller (or larger) harmonic couplings also found in the literature,  i.e.: $K \sim \, 25meV {\AA}^{-2}$ \cite{campa}, $K \sim \, 100meV {\AA}^{-2}$ \cite{barbi0},  would have the main effect in our method to shift downwards (or upwards) the denaturation temperature while leaving unaltered character of the transition and shape of the melting profiles \cite{io11a}.

Beyond the DPB Hamiltonian, Eq.~(\ref{eq:1}) introduces: \emph{ 1)} Torsional effects \cite{barbi2,io11} through the angle $\theta$ between two adjacent base pairs at the $n$ and $n-1$ sites along the molecule backbone. While the stacking is even function of $\theta$, only positive twists are allowed to describe right-handed helicoidal structures and to avoid zig-zag configurations which may also minimize the energy. \emph{2)} The solvent potential $V_{sol}(y_n)$, as proposed in Ref.\cite{collins,druk}, which adds to the Morse term thus enhancing by  $f_s D_n$ the height of the energy barrier above which the base pair dissociates.

The solvent factor $f_s$ mimics the effect of the salt concentration $[Na^+]$ on the Morse potential depth $D_n$. The latter is known to scale essentially linearly with the melting temperature $T_m$ of homogeneous GC or AT sequences (whereas, in the harmonic strong coupling limit, $T_m \propto \sqrt{D_n}$). As $T_m$ varies logarithmically with $[Na^+]$,  also the dissociation energy is assumed to increase as $D_n \propto ln[Na^+]$ \cite{santa}. Accordingly,  I take $f_s=\,0.3$ simulating a high counterion concentrations in the solvent, $[Na^+] \sim 1M$,  which screens the negatively charged phosphate groups \cite{owc,joy09}.
The length $l_s$ defines the range beyond which the Morse potential plateau is recovered and $D_n$ returns to be the fundamental energy scale.
For  $y_n > l_s$, the two strands are apart from each other and the hydrogen bond with the solvent is established. Hereafter I take $l_s=\,3 {\AA}$ whereas the possible effects due to the extension of the solvent potential range will be discussed in Section 4.3.

While Eq.~(\ref{eq:1}) refers to a system with $N + 1$ base pairs the presence of an extra base pair $y_0$ is usually handled by taking periodic boundary conditions, $y_0 = \, y_N$, which close the finite chain into a loop. This makes the model suitable for application to circular DNA although bending effects \cite{marko3,mukher} associated with tilting between base pair planes are not considered in this paper.

\subsection*{2.1 Stacking versus Twist \,}

I propose an intuitive argument, based on the principle of the energetic convenience, to support the idea that a relaxed DNA molecule modeled by Eq.~(\ref{eq:1}) should plausibly prefer twist configurations with $\theta \sim \theta_{eq}$.
To begin with, let's assume that: \emph{i)}  $N=\,100$ and \emph{ii)} in the equilibrium case, the number of base pairs per helix turn is $h=\,10$.  As $N/h$ is an integer and the helix axis lies in a plane, the strand ends line up precisely so that there is no need to introduce a curvature which would generate some \emph{writhe} \cite{white,ful1,ful2}. In this  case, the \emph{linking number} coincides with the \emph{twist}, $Lk=\,Tw$, the equilibrium twist of the relaxed configuration is $(Tw)_{eq} \equiv \,N/h=\,10$ and $\theta_{eq}=\, 2\pi /h$.

In replication and transcription, DNA unwinds thus partly dissipating the torsional stress associated with the supercoiled geometry. Double helix unwinding amounts to a twist reduction with respect to  $(Tw)_{eq}$ and generates negative supercoiling which is likely an efficient strategy to fluctuate in solution.
While \emph{supercoiling} generally refers to the molecule axis coiling upon itself, in our simple model the term indicates a variation in the helical repeat.
Thus the model in Eq.~(\ref{eq:1}) should also account for an energetic advantage of underwound configurations, $Tw < 10$,  compared to the overwound ones, $Tw > 10$. To tune the twist, imagine to cut the molecule, stretch the double helix and change the helical repeat so that $Tw$ is an integer before re-closing the ring. The DPB ladder model corresponds to the limit case, $h \rightarrow \infty$ \cite{bates}.

The intra-strand and inter-strand interactions, described by $V_S(y_n, y_{n-1})$ and $V_M(y_n)$ respectively, compete on the energy scale. If, for some displacements $y_n$, either one or the other (or both) potential term is very large the corresponding contribution to the partition function becomes vanishingly small.
$V_M(y_n)$ is in principle unbound on the $y_n < 0$ semiaxis where negative displacements depict the two complementary strands coming closer to each other than in the equilibrium state, $y_n \sim 0$. The negatively charged phosphate groups prevent however the two strands from coming \emph{too} close and the hard core of $V_M(y_n)$ accounts for such repulsive effect. Thus the base pair transverse stretchings encounter a physical cutoff $y_{min}$ which is set in our computational method \cite{io1} by the requirement $V_M(y_{min}) \sim D_n$. For AT-base pairs, it is found that $y_{min} \sim -0.2 {\AA}$ while $y_n \ll y_{min}$ would lead to inconveniently large electrostatic repulsions.
Likewise, one selects the appropriate range for GC-base pairs.
Instead, on the $y_n > 0$ semiaxis $V_M(y_n)$ is bound hence, arbitrarily large displacements may be included in the computation. Nonetheless, once the base pair dissociates, the pair mates tie to the solvent thus there is no reason to further extend the range of the stretchings. Moreover, very large $y_n$ would yield constant energy contributions to the partition function with no effect on the free energy derivatives. A cutoff  $y_{max} \sim l_s$ is then generally appropriate while, to our present purposes, even $y_{max} \sim 0.8 {\AA}$ suffices.

\begin{figure}
\includegraphics[height=5.6cm,width=6.5cm,angle=0]{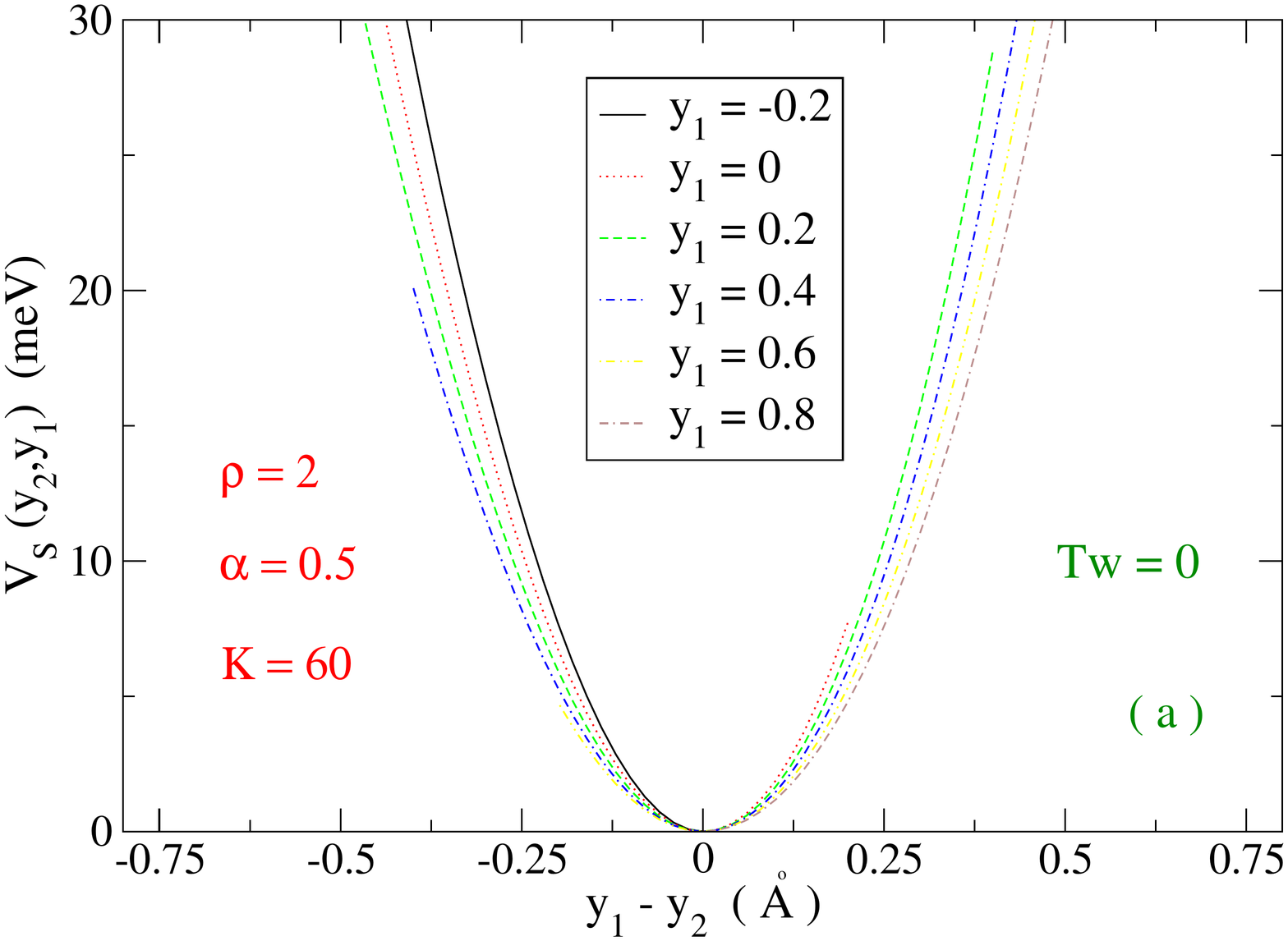}
\includegraphics[height=5.6cm,width=6.5cm,angle=0]{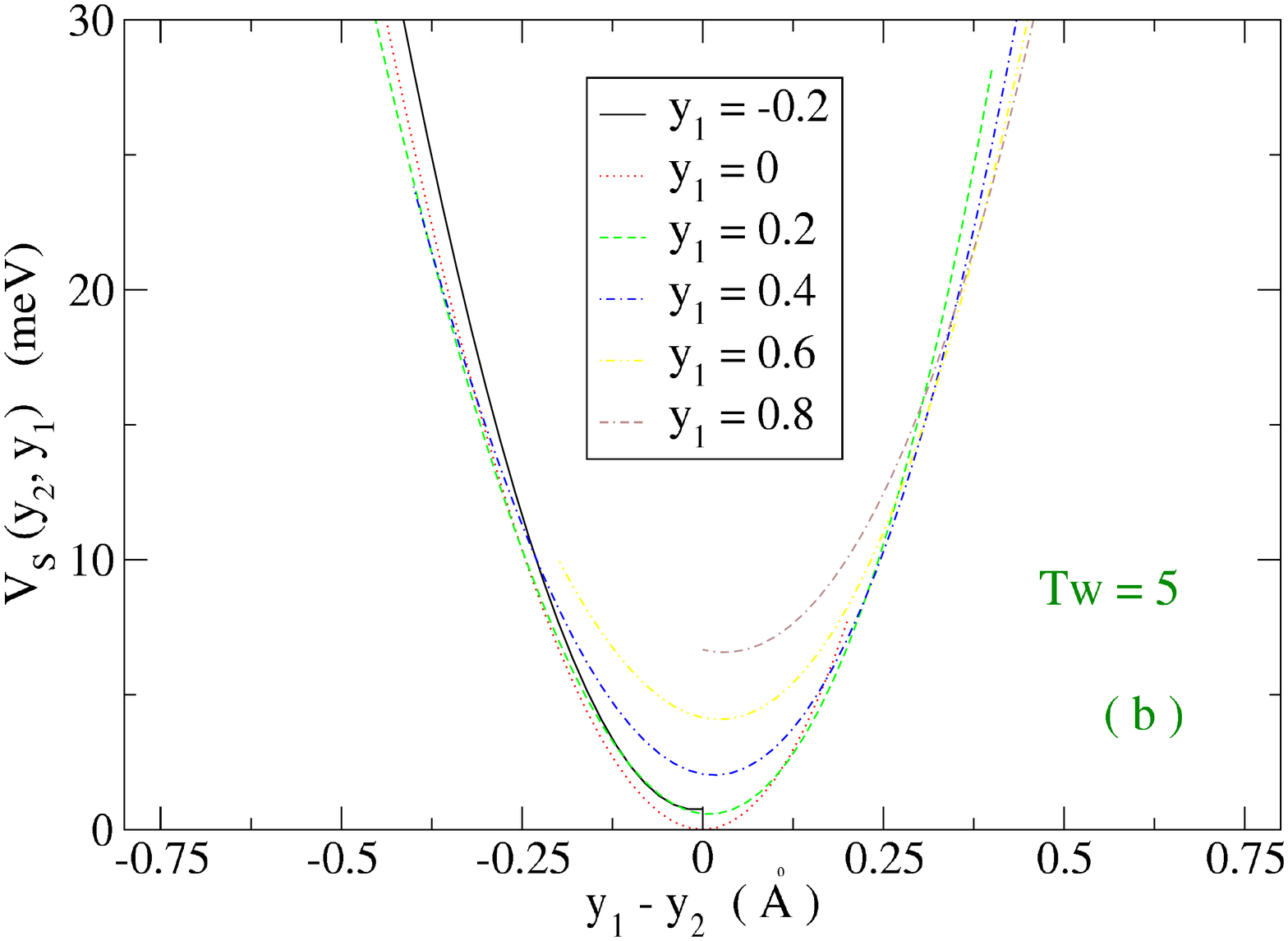}
\includegraphics[height=5.6cm,width=6.5cm,angle=0]{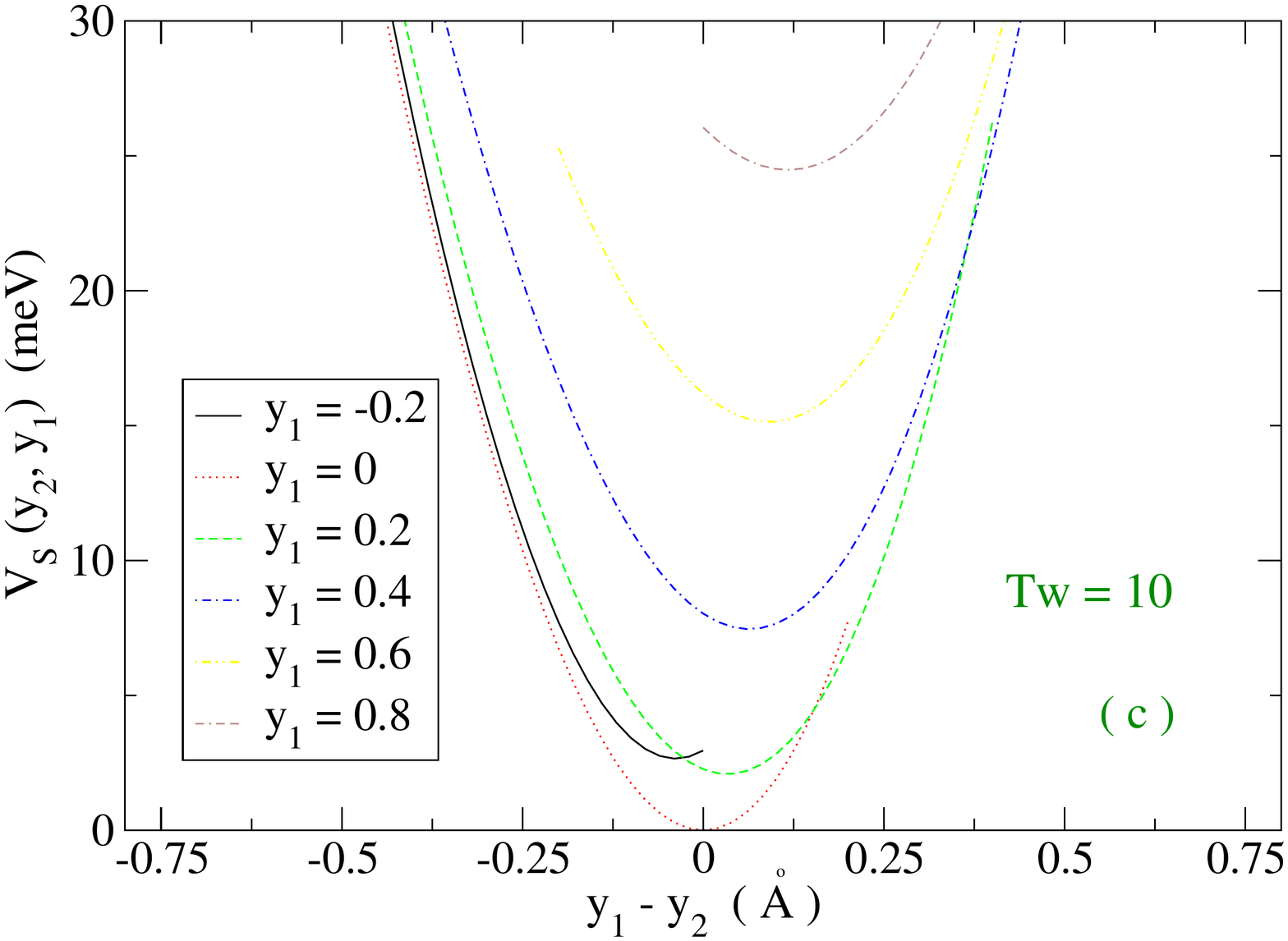}
\includegraphics[height=5.6cm,width=6.5cm,angle=0]{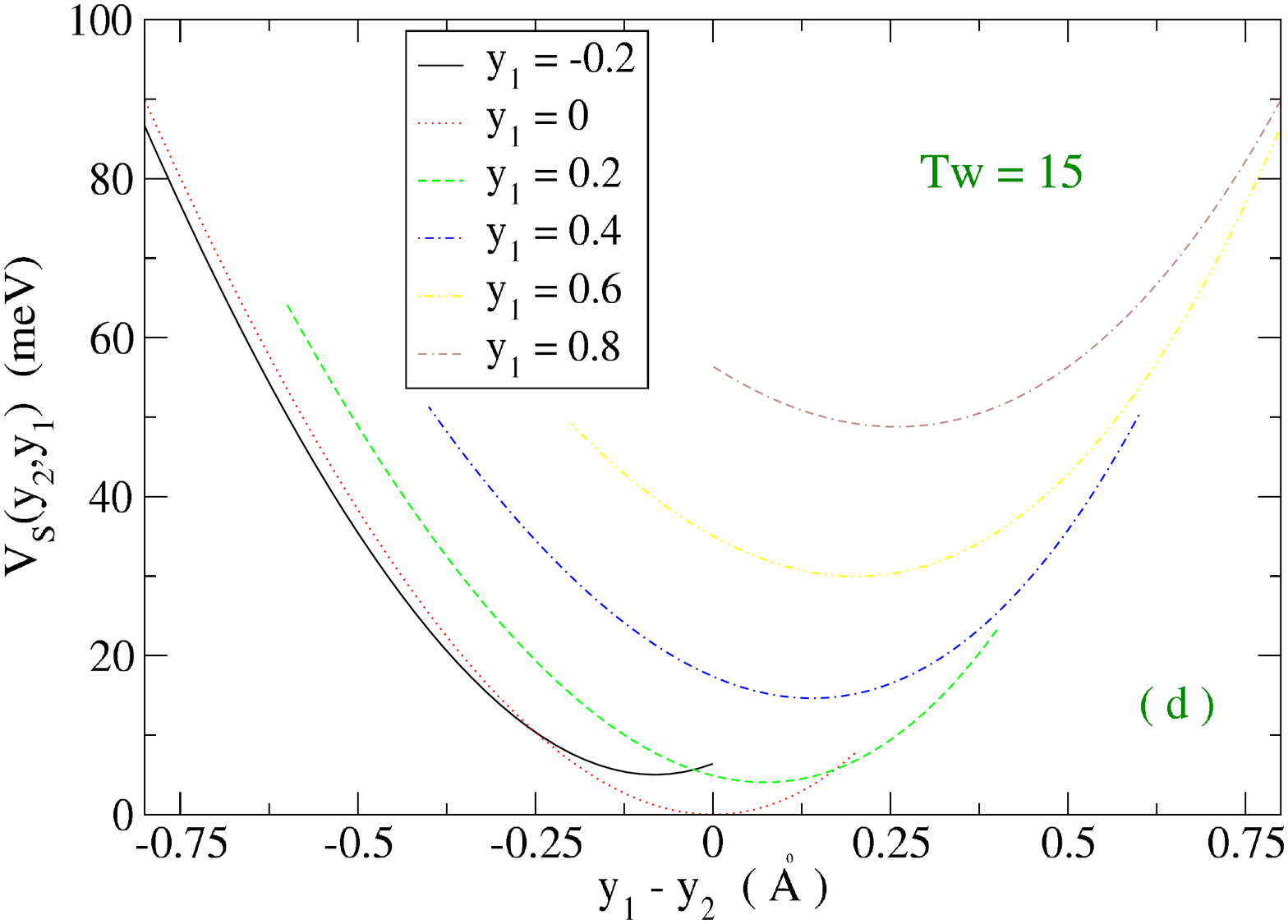}
\caption{\label{fig:1}(Color online) Stacking potential as a function of the difference between the displacements of two adjacent base pairs along the molecule backbone. Four twist conformations are assumed: (a) Dauxois-Peyrard-Bishop ladder model;  (b) Strongly underwound; (c) Equilibrium twist; (d) Strongly overwound. $\rho$ is dimensionless; $\alpha$ is in units ${\AA}^{-1}$; $K$ is in $meV K^{-1}$. }
\end{figure}

With these premises, I discuss in Fig.~\ref{fig:1} the stacking potential energetics as a function of the twist conformation tuned by the helical repeat. Two adjacent base pairs along the molecule backbone, $y_1$ and $y_2$,  are assumed to vary in the range $[-0.2,\,0.8]{\AA}$ and the relative stacking  $V_S(y_2, y_{1})$ is plotted versus the difference  $y_1 - y_2$  pinning $y_1$ to six values in that range.

Fig.~\ref{fig:1}(a) refers to the DPB ladder model, $\theta=\,0$ in Eq.~(\ref{eq:1}). Whatever value is assigned to $y_1$, the second base pair can adjust its position so as to keep $V_S$ low in energy: i.e., setting $y_1=\,0.8 {\AA}$ and varying $y_2$, one finds a range $\,y_1 - y_2 \in [0, 0.5]{\AA}$ in which $V_S$ does not exceed $\sim 30meV$, the Morse plateau for AT-pairs. Assuming $y_1=\,0.4 {\AA}$,  $V_S < 30meV$ in the broad range $y_1 - y_2 \in [-0.4, 0.4]{\AA}$.  Likewise for any other $y_1$. Obviously the  potential minimum always occurs for $y_2 =\, y_1$ but the stacking energies remain generally of order of $D_{AT}$ even in the case of large differences $y_1 - y_2$.    Because of a fluctuation, a base pair may get a significant displacement ($\sim 0.8{\AA}$) out of the equilibrium nevertheless the adjacent base pair can move quite freely keeping the stacking potential at values comparable with $V_M$. Thus the stacking of the DPB model gives little stability to the double strand molecule as it does not \emph{sufficiently discourage} large relative fluctuations which therefore contribute to the partition function.
As a general requirement the model should be both flexible enough to absorb fluctuations and stacked enough to guarantee some stability against thermal disruption.

This goal is achieved by introducing a twist through a finite $\theta$ as shown in Fig.~\ref{fig:1}(b) and, thoroughly, in Fig.~\ref{fig:1}(c) which displays the equilibrium conformation. Mostly in the latter, according to the value assigned to $y_1$, some $y_2$ displacements are discouraged as they appear too costly on the energy scale. If  $y_1 \sim 0.8 {\AA}$, $y_2$ can only assume large enough values  (such that $y_1 - y_2 \in [0, 0.3]{\AA}$) to minimize the stacking energy but, even in this restricted range, $V_S$ is much larger than in Fig.~\ref{fig:1}(a) signalling that the torsion is indeed effective in providing some restoring force to the molecule.
The energy spacings among the minima in Fig.~\ref{fig:1}(c) indicate that some couples of base pairs ($y_1$, $y_2$) are less likely to contribute to the partition function and, at the same time, they also show the system capability in accepting moderately large fluctuations. Taking $y_1 =\, 0.4 {\AA}$, the possible displacements for $y_2$ which cost less than $30meV$ are such that $y_1 - y_2 \in [\sim -0.35, 0.4]{\AA}$ hence, $y_2$ has a broad range to explore and can become even larger than $y_1$. Unlike Fig.~\ref{fig:1}(a), the potential minimum does not correspond to the translation $y_2 =\,y_1$ but rather to a slightly positive $y_1 - y_2$. $y_2$ follows $y_1$ to pull it back. This is the restoring effect yielding flexibility to the system.

One may wonder whether larger twists, overwound conformations, might become even more energetically convenient than $Tw=\,10$. The answer is however negative as shown in Fig.~\ref{fig:1}(d) where $V_S$ is plotted for the case $Tw=\,15$. Now the energy scale changes drastically and the energy cost to follow a sizeable fluctuation in $y_1$ becomes too high:  if  $y_1 \geq 0.4 {\AA}$, there is no $y_2$ value to keep $V_S$ comparable to the Morse dissociation energy. Only taking $y_1$ in the range  $[-0.2, 0.2]{\AA}$ there exists a likewise small $y_2\,$ range  such that $V_S \leq D_{AT}$ for  $y_1 - y_2 \in [\sim -0.25, 0.25]{\AA}$. This means physically that only rigid structures are allowed and most of the base pair fluctuations are discarded as they overcome the Morse plateau  thus yielding too large contributions to the classical action.
Note that the number of turns added in Fig.~\ref{fig:1}(d) (with respect to $Tw=\,10$) is equal to the number of turns subtracted in Fig.~\ref{fig:1}(b). Nevertheless an interesting, substantial asymmetry is found between the strongly overwound (Fig.~\ref{fig:1}(d))and the strongly  underwound (Fig.~\ref{fig:1}(b)) conformation: the latter displays an energy scale much closer than the former to the equilibrium case (Fig.~\ref{fig:1}(c)).

Summing up, the double helix appears to be shaped by the balance between two competing needs: allowing for those  fluctuational openings which are vital to biological functions meanwhile forbidding those fluctuations which may become too large and energetically costly. General considerations based on the mesoscopic Hamiltonian in Eq.~(\ref{eq:1}) suggest that twists conformations with $Tw \sim \,10$ emerge as the most suitable for the relaxed molecule.
While the argument proposed so far cannot discern any appreciable difference between slightly underwound  ($Tw=\,9$) and slightly overwound  ($Tw=\,11$) conformations, an energetic advantage is expected for stronger underwound ($Tw < \,9$) molecules over the stronger overwound ($Tw > \,11$) ones.
After establishing that the stacking potential in Eq.~(\ref{eq:1}) captures the essentials of the interactions in the helicoidal geometry I proceed to develop a more quantitative analysis in terms of the path integral formalism \cite{fey} which also incorporates the environmental conditions due to temperature and solvent.

\section*{3. Path Integral  }

The double helix base pairs permanently fluctuate around their equilibrium positions represented by $y_n \sim 0$ in Eq.~(\ref{eq:1}). Fluctuations get larger by increasing temperature and produce temporary openings, bubbles, which are peculiar of the biological functions \cite{ares1,bonnet,metz06}. Such fluctuational openings initiate at specific sites and propagate along the molecule backbone as a consequence of cooperative effects in the stacking.

In an effort to account for the temperature dependent DNA dynamics, I have proposed \cite{io} to apply path integral techniques to the mesoscopic Hamiltonian in Eq.~(\ref{eq:1}). This permits to investigate the sequence specificity in the molecule and also to model the base pair fluctuational effects in the computation of the partition function. To accomplish a finite temperature description, the appropriate theoretical tool is the \emph{imaginary time} path integral method \cite{fehi} where the imaginary time $\tau$, an analytic continuation of the real time, runs on a scale which is set by the inverse temperature: $\tau \in [0,\,\beta]$ with $\beta=\, (k_B T)^{-1}$. The length $\beta$ is partitioned in $N_\tau$ intervals each identified by a specific $\tau_i$.  As this partitioning is repeated at any temperature while $N_\tau$ is kept constant, each $\tau_i$ can vary inside the range whose upper bound is $\beta$.
Accordingly the Euclidean action integral transforms into a discrete sum (over $i$) of Hamiltonian contributions (evaluated at different $\tau_i$).

Thus, the space-time mapping technique for the model in Eq.~(\ref{eq:1}) assumes that the $N$ base pair radial displacements $y_n$ are thought of as one dimensional paths $x(\tau_i)$ which are periodic functions of $\tau_i$, $x(\tau_i)=\,x(\tau_i + \beta)$.  $i$ numbers the base pairs along the $\tau$-axis and $\tau_i \in [0,\,\beta]$. As $x(0)=\,x(\beta)$, a molecule configuration is given by $N_\tau=\,N$ paths and, in the discrete imaginary time lattice, the separation between nearest neighbors base pairs is $\Delta \tau =\,\beta / N_\tau$.

Then, the displacements in Eq.~(\ref{eq:1}) transform as:

\begin{eqnarray}
& &y_n \rightarrow x(\tau_i) \,  \nonumber
\\
& &y_{n-1} \rightarrow x(\tau_i - \Delta \tau)\,
\label{eq:3}
\end{eqnarray}

The base pair paths are written in Fourier series with cutoff $M_F$

\begin{eqnarray}
& &x(\tau_i)=\, x_0 + \sum_{m=1}^{M_F}\Bigl[a_m \cos(\omega_m \tau_i) + b_m \sin(\omega_m \tau_i) \Bigr] \, \nonumber
\\ \,
& &\omega_m =\, {{2 m \pi} / {\beta}}\, .
\label{eq:3a}
\end{eqnarray}

Accordingly, a set of coefficients $\{x_0 , a_m , b_m\}$ univocally defines a molecule configuration at a specific $T$, with the $N$ base pairs given by $\{x(\tau_i), \, i =\,1\,,..,\,N_\tau \}$.
As a molecule state is defined by a point in the paths configuration space, path integration amounts to sum over a large ensemble of distinct molecule states each weighted by a Boltzmann factor. The method fully incorporates fluctuational effects around the ground state and monitors the bubble formation around and above room temperature. The classical partition function for the system in Eq.~(\ref{eq:1}) reads

\begin{eqnarray}
& &Z_C=\,\oint \mathfrak{D}x\exp\bigl[- \beta A_C\{x\}\bigr]\, \nonumber
\\
& &A_C\{x\}=\, \sum_{i=\,1}^{N_\tau} \Bigl[{\mu \over 2}\dot{x}(\tau_i)^2 + V_S(x(\tau_i),x(\tau_i - \Delta \tau)) + V_M(x(\tau_i)) + V_{sol}(x(\tau_i)) \Bigr] \, \nonumber
\\
\label{eq:3b}
\end{eqnarray}

where the measure of integration $\mathfrak{D}x$ is expressed in terms of the Fourier coefficients by

\begin{eqnarray}
& &\oint \mathfrak{D}x\equiv {1 \over {\sqrt{2}\lambda_\mu}}\int_{-\Lambda^0_T}^{\Lambda^0_T} dx_0 \prod_{m=1}^{M_F}\Bigl({{m \pi} \over {\lambda_\mu}}\Bigr)^2 \times \int_{-\Lambda_T}^{\Lambda_T} da_m \int_{-\Lambda_T}^{\Lambda_T} db_m \, \, . \,
\label{eq:3c}
\end{eqnarray}

Defining the free energy, $F=\,\beta^{-1}\ln Z_C$, and the entropy, $S=\,- dF/dT$, the thermodynamical properties of the system can be computed via Eqs.~(\ref{eq:3b}), ~(\ref{eq:3c}) taking a temperature range $[T_{min}, T_{max}]$ in which fluctuational openings are expected to occur. The bounds, $T_{min}=\,260K$ and $T_{max}=\,390K$, are chosen hereafter.

Say $N_{eff}$ the number of integration points in Eq.~(\ref{eq:3c}) to be selected on the base of thermodynamical and model potential physical requirements as detailed in Refs.\cite{io,io1}. In particular, after setting an initial $N_{eff}$ value at $T_{min}$ which ensures numerical convergence for $Z_C$, the programme evaluates the free energy derivatives at any $T > T_{min}$. If, for a given $T$, the entropy derivative is not positive then $N_{eff}$ is increased and the thermodynamics is re-calculated. The procedure is reiterated for any $T$ until the second law of thermodynamics is fulfilled throughout the whole temperature range.
Eventually the method yields a $N_{eff}(T)$ plot which denotes the  number of possible trajectories for the $i-th$ base pair whereas $N_\tau \times N_{eff}$ represents the total number of paths contributing to $Z_C$. $N_\tau \times N_{eff}$ yields the measure of the path ensemble size which, consistently with the entropy law, turns out to be a growing function of $T$.

${\lambda_\mu}=\,\sqrt{{\pi } / {\beta K}}$, is the thermal wavelength in the classical regime.
The cutoffs ${\Lambda^0_T}$ and ${\Lambda_T}$ in Eq.~(\ref{eq:3c}) remind us that the path amplitudes are temperature dependent. They are derived analytically noticing that the integration measure normalizes the free particle action \cite{kleinert}:

\begin{eqnarray}
\oint \mathfrak{D}x(\tau)\exp\Bigl[- \int_0^\beta d\tau {\mu \over 2}\dot{x}(\tau)^2  \Bigr] =\,1.
\label{eq:4} \,
\end{eqnarray}

Hence, inserting Eqs.~(\ref{eq:3a}), ~(\ref{eq:3c}), the l.h.s. of Eq.~(\ref{eq:4}) transforms into a product of Gau{\ss}ian integrals whose normalization conditions yield:

\begin{eqnarray}
& &{\Lambda^0_T}=\,\lambda_\mu/\sqrt{2} \, \nonumber
\\
& &\Lambda_T =\,{{U \lambda_\mu}  \over {m \pi^{3/2}}}.
\,
\label{eq:5}
\end{eqnarray}

Thus the temperature effects in the molecule states ensemble is included in the computation also through the path amplitude cutoffs with $\propto \sqrt{T}$ dependence. The dimensionless $U$ can be tuned to check the relevance of cutoff effects on the system thermodynamics \cite{zhang}.

\section*{4. Results }

The model is applied to a fragment with $N_\tau =\,100$ base pairs. The first $\tau_{i=\,1},...,\tau_{i=\,48}$ sites from left have the same sequence as the \emph{L48AS }fragment of Refs.\cite{zocchi2} which shows fluctuations in the AT-rich side triggering the opening of distant GC pairs. In fact the melting curves as obtained by UV absorption measurements show that the base pairs open essentially in two stages, at two distinct temperatures. However, the fraction of base pairs opening in the first stage is larger than the whole fraction of AT- base pairs in the sequence. This implies that also GC- base pairs feel the effects of neighboring AT-rich regions, in particular of TATA boxes.

These long range effects make the sequence of interest to us. Next, $52$ AT-base pairs are added to \emph{L48AS} on the right side to further develop fluctuational openings \cite{hwa}. While the fragment has been studied in previous works \cite{io11a,io11}, it is here reconsidered in order to analyze the anharmonic effects in twisted geometries.

The $\tau_{i=\,101}$ site closes the double helix into a loop and therefore hosts a GC base pair. Inside AT-rich regions, pyrimidine-purine stacking steps favor local unwinding hence TA/TA steps are more efficient than AA/TT in driving bubble formation \cite{yakov,metz06,metz09}.  The whole double strand sequence is:

\begin{eqnarray}
& &GC + 6AT +  GC + 13AT + 8GC + AT + 4GC + \nonumber
\\
& &AT + 4GC + AT + 8GC + [49-100]AT + GC \, . \,\nonumber
\\
\label{eq:6}
\end{eqnarray}

\subsection*{4.1 Melting Profiles \,}

As the UV signal changes quite abruptly when base pairs dissociate, the fraction of open base pairs is generally defined in terms of the Heaviside step function $\vartheta(\bullet)$ as:

\begin{eqnarray}
& &f =\, {1 \over {N_\tau}}\sum_{i=1}^{N_\tau} \vartheta\bigl(< x(\tau_i) > - \zeta \bigr) \, , \,
\label{eq:7}
\end{eqnarray}

where $< x(\tau_i) >$, the ensemble average for the displacement of the $i-th$ base pair, is computed via Eq.~(\ref{eq:3b}) with potential parameters as above. The {\it threshold} $\zeta$ yields a criterion to establish whether an average base pair displacement is open, $<x(\tau_i)>\, \geq\, \zeta$, or not.

It should be remarked that Eq.~(\ref{eq:7}) does not define univocally, on the $T$ axis, at which stage the strand separation is large enough to produce full denaturation but, it provides a tool to evaluate quantitatively the amplitude of the fluctuational openings and the number of base pairs which participate in the bubble formation.
While the choice of $\zeta$ is somewhat arbitrary, in view of the fact that the B-DNA equilibrium diameter is $\sim 20{\AA}$, by taking $\zeta$ in the range $\sim [0.5 - 2]{\AA}$ \cite{pey4,ares,pey9} one confidently explores the formation of nonlinear fluctuational openings in the appropriate temperature domain.

\begin{figure}
\includegraphics[height=6.0cm,angle=0]{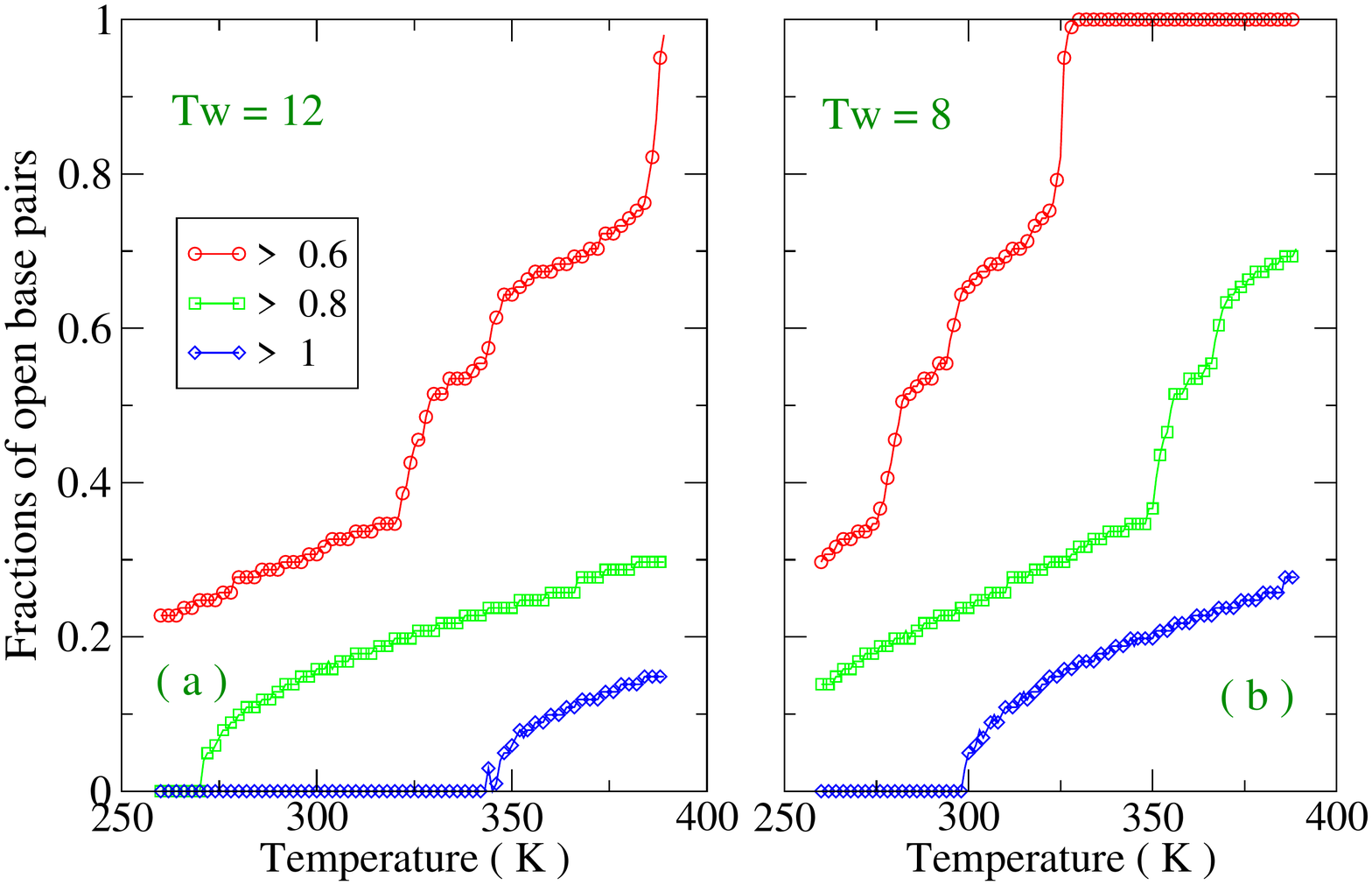}
\includegraphics[height=6.0cm,angle=0]{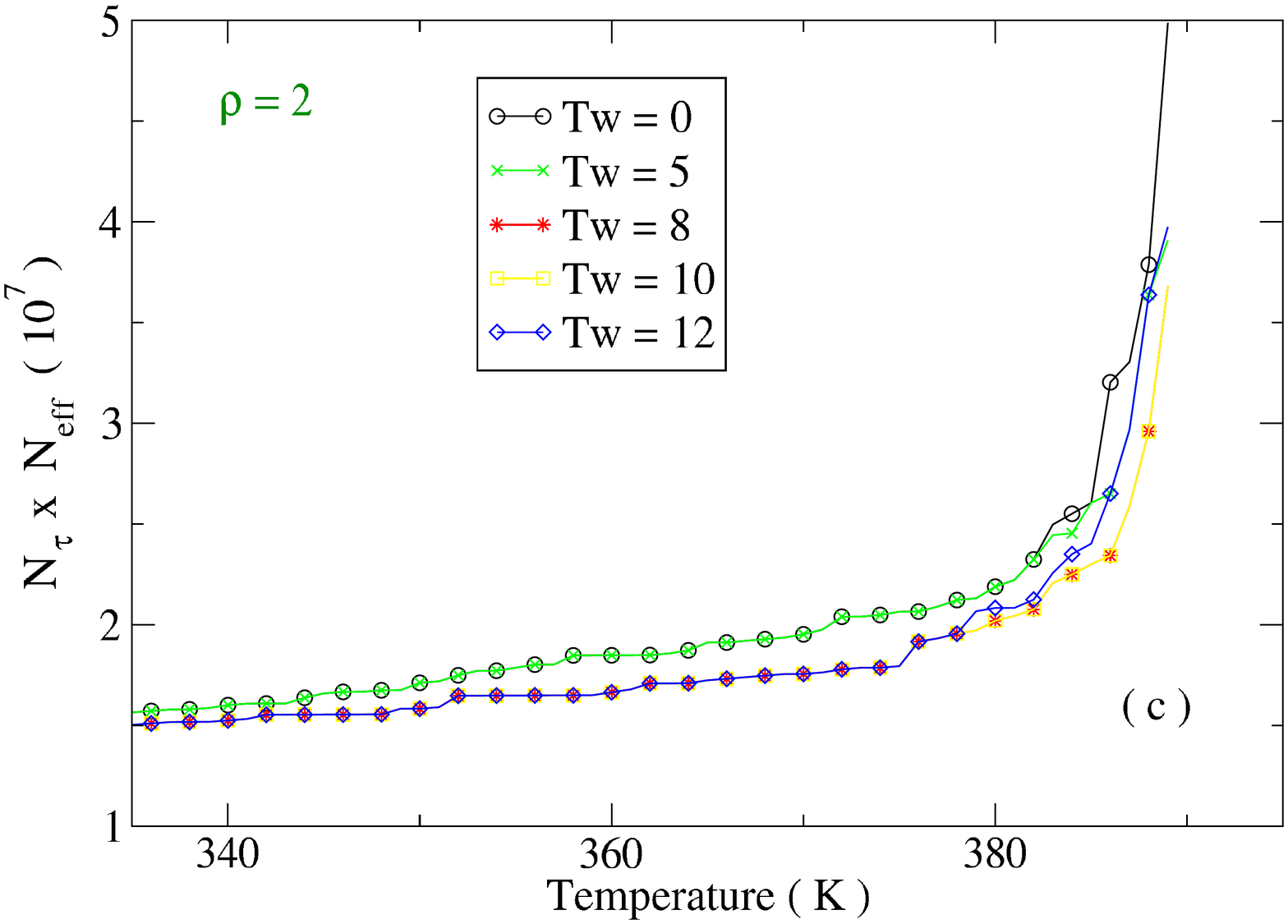}
\vspace*{.05in}
\caption{\label{fig:2}(Color online) Fractions of base pairs larger than $0.6, \,0.8, \, 1 {\AA}$ versus temperature in {(a)} moderately overwound,   {(b)} moderately underwound double helix.  {(c)} Total number of paths contributing to the partition function for four twist (and the ladder) conformations. The model potential parameters are as in Fig.~\ref{fig:1}. }
\end{figure}

The melting profiles given by Eq.~(\ref{eq:7}) are plotted in Fig.~\ref{fig:2} for moderately overtwisted and untwisted conformations: two turns of twist more (in Fig.~\ref{fig:2}(a)) and less (in Fig.~\ref{fig:2}(b)) respectively than $(Tw)_{eq}$.  Three $\zeta$ values are chosen to calculate $f$ as a function of $T$ in the denaturation range. There is a remarkable effect due to the twist. For $Tw=\,8$,  all average displacements ($f = \,1$) get larger than $\zeta=\,0.6{\AA}$ at $T \sim \,325K$ and $\sim\, 30 \%$ ($f \sim \,0.3$) get larger than $\zeta=\,1{\AA}$ in the upper $T$ range. For $Tw=\,12$,  $f \sim \,1$ for $\zeta=\,0.6{\AA}$ only at temperatures around $400K$ whereas, in such range, only $\sim\, 15 \%$ of the average displacements exceed the $\zeta=\,1{\AA}$ threshold. Overtwisting confers rigidity to the molecule while moderate untwisting provides flexibility.

In both cases however segments of closed base pairs persist well inside the denaturation regime as indicated by recent neutron scattering measurements on B-DNA \cite{theo2}.
On the other hand, the $Tw=\,8$ conformation shows that also the GC-base pairs ($26 \%$ of the total) exceed the threshold $\zeta=\,0.6{\AA}$ at $T \sim \,325K$. In particular, the second GC-pair from left (in Eq.~(\ref{eq:6})) has a large $T$ dependent average displacement being embedded in a broad AT-rich region. This is consistent with the experimental findings on \emph{L48AS} cited above.

In Fig.~\ref{fig:2}(b), three main sharp enhancements appear at $T \sim \, 280K$, $T \sim \, 325K$ ( $\zeta=\,0.6{\AA}$) and $T \sim \, 350K$ ( $\zeta=\,0.8{\AA}$ ). Two smaller steps are at $T \sim \, 300K$ ( $\zeta=\,0.6, \,1{\AA}$ plots).
These are signatures of multistep openings around and above room temperature but $\sim\, 70 \%$ of the average displacements are still smaller than $\zeta=\,1{\AA}$ at $T \sim \, 400K$ suggesting that moderate untwisting is consistent with an overall thermal stability of the molecule.

The $Tw=\,8$ conformation is then expected to be similar, in terms of energetic convenience, to the $(Tw)_{eq}$ case.
This is indeed confirmed by Fig.~\ref{fig:2}(c) where $N_\tau \times N_{eff}$ is plotted in the high $T$ range.

As explained in Section 3, $N_\tau \times N_{eff}$ is the number of possible paths which have been selected to describe the base pair displacements. These paths are included in the computation of $Z_C$. While $N_\tau \times N_{eff}$ varies with temperature, for a given $T$, it also depends on the twist geometry.
The lower the number of paths required in the computation of the thermodynamical properties, the higher is the stability of the twist conformation. $N_\tau \times N_{eff}$ is in fact a measure of the system capability to sustain thermal fluctuations allowing for local openings without fully denaturating. The computation includes ensembles of order $10^7$ paths. The DPB ladder model and the strongly untwisted ($Tw=\,5$) conformation display the largest $N_\tau \times N_{eff}$ pointing to a major instability degree whereas relaxed ($Tw=\,10$) and moderately untwisted ($Tw=\,8$) conformations are the stablest ones. In fact the latter two plots overlap. Lying between the unstable and stable conformations, the overtwisted ($Tw=\,12$) molecule shows a larger (than the $Tw=\,8, 10$ cases) gradient in the upper $T$ range.

In summary, the analysis of the path ensemble size as a function of twist shows that a moderate unwinding of the double helix provides overall thermal stability to the molecule consistently with the computation of the average base pair displacements which determine the melting profiles.

\subsection*{4.2 Anharmonic Stacking \,}

\begin{figure}
\includegraphics[height=6.0cm,width=8.5cm,angle=0]{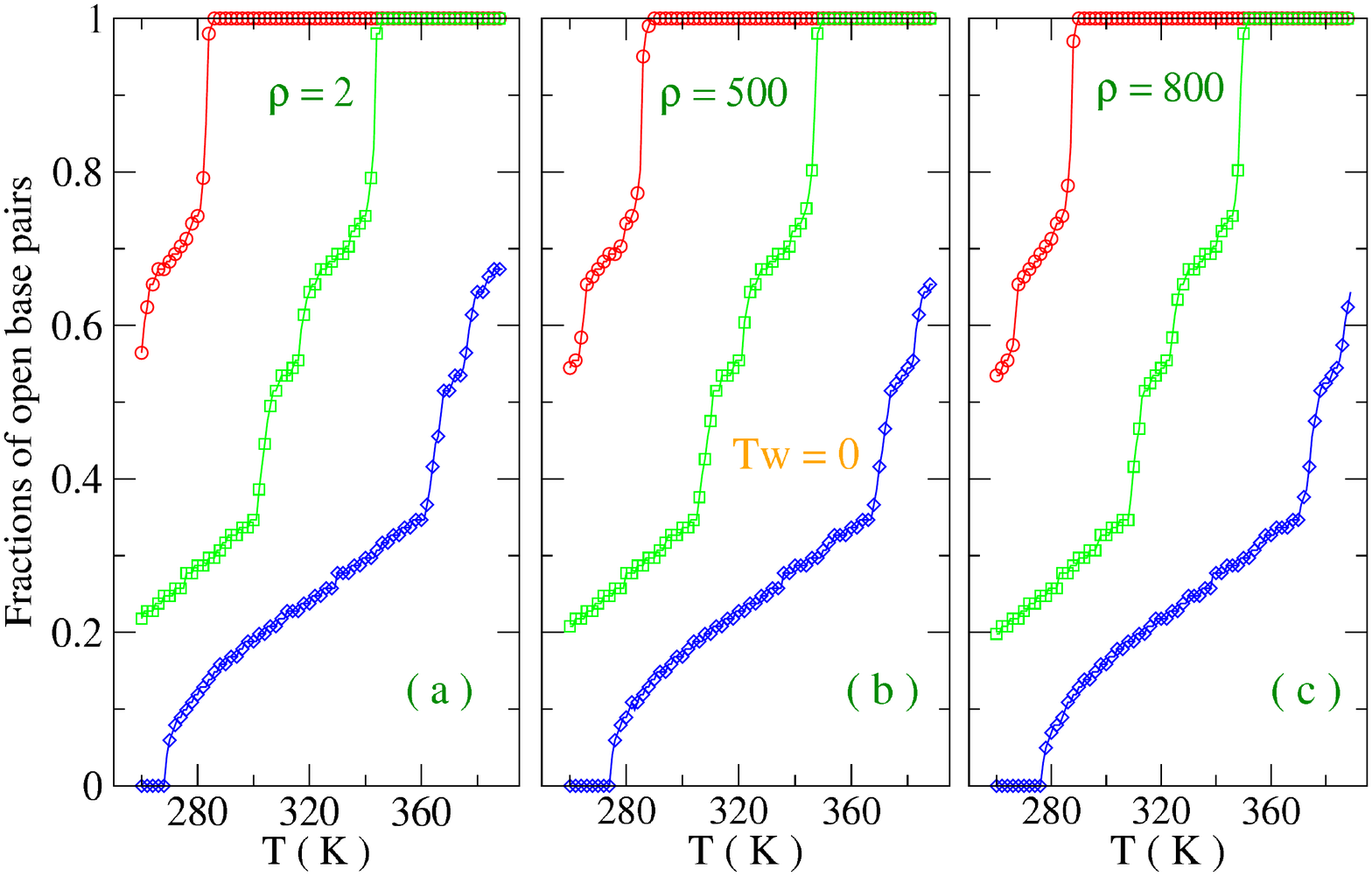}
\caption{\label{fig:3}(Color online) Fractions of average displacements larger than $0.6 \, (\bigcirc),  \,0.8\, (\square), \, 1\, (\lozenge) {\AA}$ versus temperature in the  DPB ladder model. Three anharmonic stacking $\rho$ are assumed. The other stacking potential parameters are as in Fig.~\ref{fig:1}. }
\end{figure}

I investigate the interplay between twisting and stacking anharmonicity also in view of the special role given to the latter in the DPB model where a finite $\rho$  induces a sharp denaturation transition driven by sizeable melting entropy \cite{pey2}.
Taking a ratio $\alpha /a_n \sim 0.1$, the range of the entropic barrier is here assumed to be longer than that of the Morse potential \cite{cocco}.
In Fig.~\ref{fig:3} the melting profiles of the DPB ladder model are calculated by increasing $\rho$ over the previously used value, $\rho=\,2$. Even very large $\rho$ produce scant variations in the denaturation patterns pointing to a substantial irrelevance of the $\rho$ driven anharmonicity for this conformation.
Comparing Fig.~\ref{fig:3}(a) to Figs.~\ref{fig:2}(a),(b), it is seen that the base pair opening in the ladder model is shifted at lower $T$, with $f \sim \,1$ (for $\zeta=\,0.8{\AA}$) already at $T \sim \,280K$. This confirms that, in the absence of twist, the model is largely unstable consistently with the $Tw=\,0$ plot in Fig.~\ref{fig:2}(c) and with the expectations provided by Fig.~\ref{fig:1}(a).

\begin{figure}
\includegraphics[height=6.0cm,width=8.5cm,angle=0]{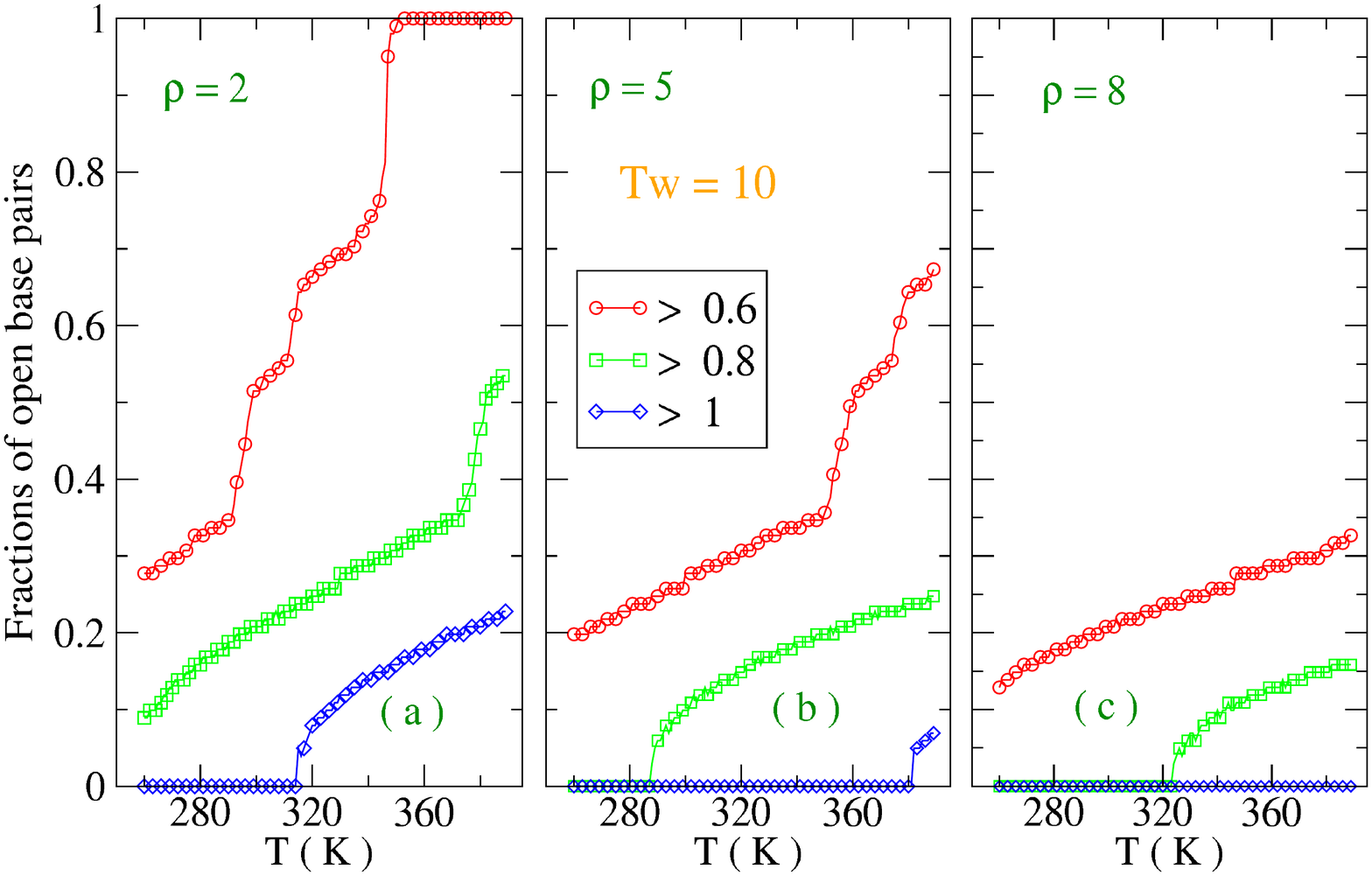}
\caption{\label{fig:4}(Color online) As in Fig.~\ref{fig:3} but for the equilibrium twist conformation. In {(b)} and {(c)}, $\rho$ is smaller by a factor $100$ than in  Fig.~\ref{fig:3}(b),(c) respectively. }
\end{figure}

Quite different is the physical picture emerging from Fig.~\ref{fig:4} where the efficacy of the stacking anharmonicity is sampled on the $(Tw)_{eq}$ conformation. Here, even slight enhancements over the $\rho=\,2$ value \emph{ i)} shift upwards along $T$ the opening of the average base pair displacements and \emph{ii)} flatten the melting profiles suggesting that the denaturation becomes more gradual.  Note that $\rho$ values in Figs.~\ref{fig:4}(b),(c) are two orders of magnitude smaller than in Figs.~\ref{fig:3}(b),(c) respectively. The $\rho=\,8$ plot says that, even at $T \sim \,400K$, there are no average displacements larger than $1 {\AA}$ while only $35\%$ are larger than $0.6 {\AA}$. Thus it is the twisting that switches on the $\rho$ effect in the model. Anharmonic stacking induces those cooperative interactions  along the molecule backbone which are peculiar of the fluctuational openings \cite{io}. At the same time anharmonic stacking renders the double helix flexible hence it does increase the molecule resilience against the whole thermal disruption of the hydrogen bonds.
In this sense, anharmonicity is a stabilizing factor for the double helix.

\subsection*{4.3 Solvent Potential \,}

\begin{figure}
\includegraphics[height=6.0cm,width=7.0cm,angle=0]{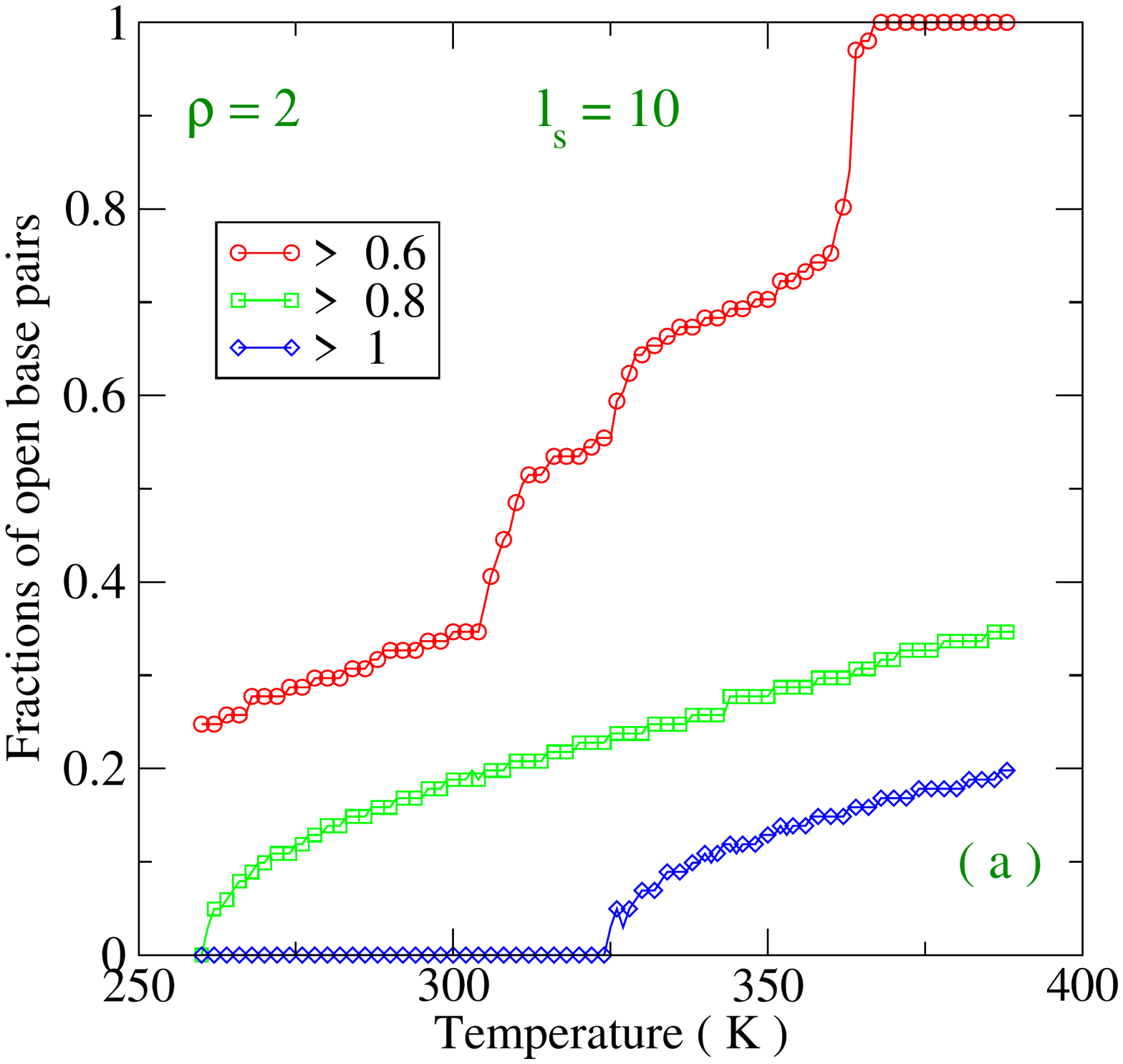}
\includegraphics[height=6.0cm,width=7.0cm,angle=0]{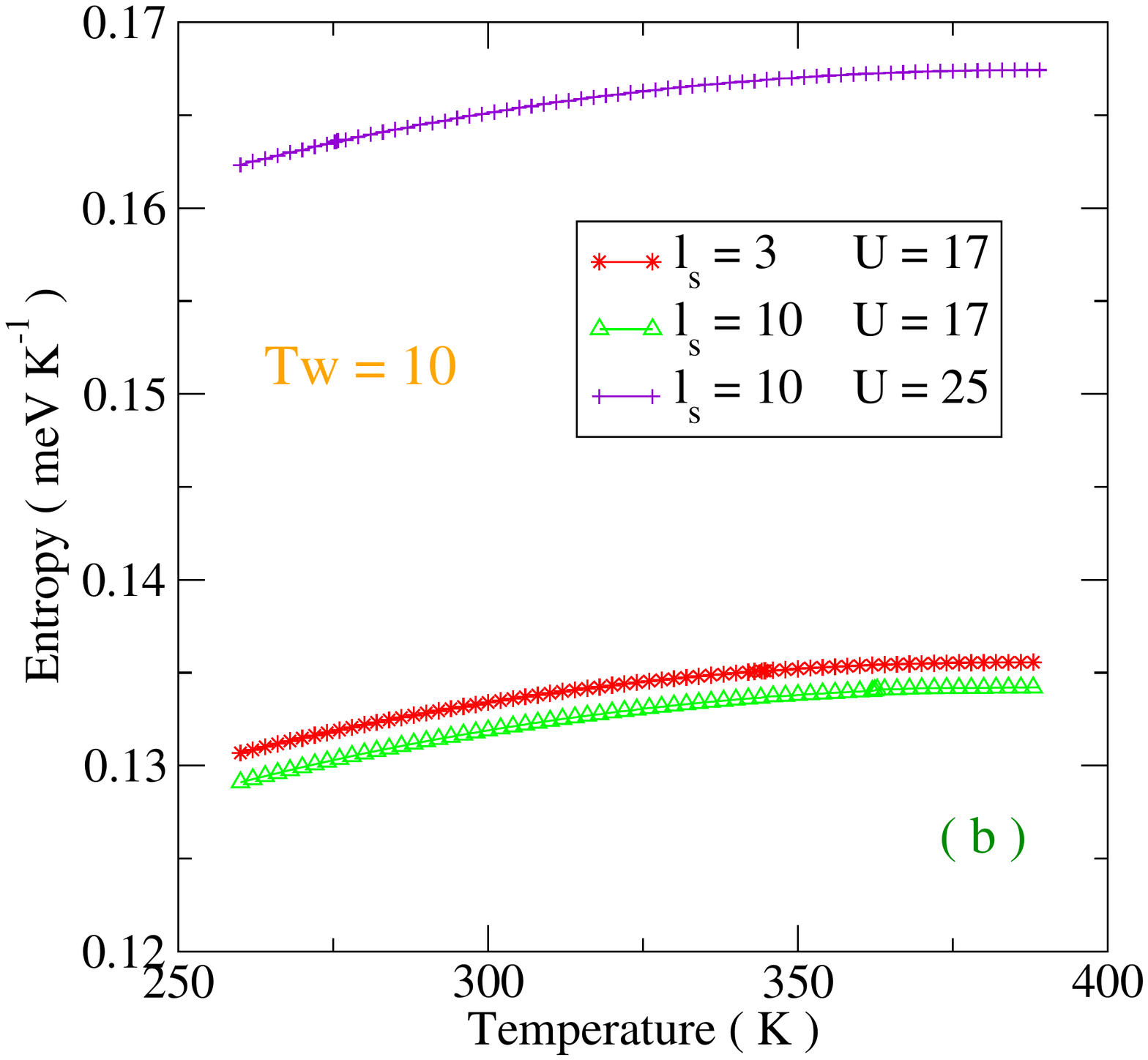}
\caption{\label{fig:5}(Color online) (a) Fractions of average displacements larger than $0.6, \,0.8, \, 1 {\AA}$ in the equilibrium twist configuration for a large solvent potential range, $l_s=\,10{\AA}$. The stacking potential parameters are as in Fig.~\ref{fig:3}{(a)}.  {(b)} Equilibrium twist entropy as a function of \emph{i)} $l_s$ and \emph{ii)} the dimensionless path amplitude cutoff $U$. }
\end{figure}

Finally I discuss the effect of the solvent potential in Eq.~(\ref{eq:1}) extending the range well beyond the length $l_s=\,3\,{\AA}$ used so far. $l_s$ defines the spatial range within which the solvent modifies the plateau of the Morse potential.
A study of the DPB model \cite{weber} with the same potential term as in Eq.~(\ref{eq:1}), concludes that growing $l_s$ would cause a sharp, first order-like, melting transition which is however smoothed by inclusion of a small twist angle. The melting profiles and the entropy plots shown in Fig.~\ref{fig:5} do not support such conclusions.   Broadening the potential range shifts the denaturation steps upwards (Fig.~\ref{fig:5}(a)) but it does not change the smooth character of the denaturation transition which is moreover independent of the torsional degree: whatever twist angle is chosen the entropy grows continuously.   The curves for the equilibrium twist conformation are given in Fig.~\ref{fig:5}(b) both for $l_s=\,3{\AA}$ and $l_s=\,10{\AA}$. In the latter case the entropy is slightly reduced consistently with the physical expectations attributing  an overall stabilizing role to the solvent \cite{blake}.
Instead, the entropy grows by enhancing the cutoff $U$ in Eq.~(\ref{eq:5}) due to the fact that much larger path amplitudes are included in the computation but even this does not sharpen the denaturation which remains continuous, second order-like.  At the denaturation steps the melting entropy, $\sim 10^{-4} meV K^{-1}$, is smaller by a factor $100$ than that found in the DPB ladder model by transfer integral method \cite{theo} albeit for homogeneous DNA  \cite{cule}.

\section*{5. Discussion }

The path integral formalism has been applied to a circular heterogeneous DNA molecule modeled by a generalized Dauxois-Peyrard-Bishop Hamiltonian which incorporates a solvent potential and accounts for the helicoidal geometry. Inter-strand and intra-strand interactions are nonlinear. Sampling about $10^7$ molecule states in the path configuration space, I have studied the system thermodynamics focussing on the interplay between stacking anharmonicity and DNA supercoiling. The latter is tuned by varying the helical repeat with respect to the equilibrium conformation of room temperature B-DNA. Twisting has the effect to bring closer to each other non-adjacent bases in the stack thus accounting for long range interactions which shape the molecule dynamics.
Computations of melting profiles and size of the path ensemble  consistently indicate that moderately untwisted helices, as much as the equilibrium conformation, are energetically favored with respect to the overtwisted ones. Then the path integral analysis quantitatively confirms the predictions based on the energetics of the stacking potential model versus degree of twist.

At variance with previous studies, it is found that anharmonic stacking has little weight in the DPB model with zero twist whereas it becomes effective in helicoidal geometries and mainly for twist angles around the equilibrium conformation. In the latter an enhanced anharmonicity smooths the melting profiles and shows an overall stabilizing role.
Thus, I put forward a picture in which the nonlinear stacking potential is essential to make the molecule flexible and therefore able to sustain those thermal fluctuational openings which are key to the biological functioning.
This points to an interesting relation between intrinsic nonlinear character of the microscopic interactions and molecule topology. Accordingly anharmonicity is negligible in the ladder conformation which can be easily disordered by disruption of the hydrogen bonds.
The calculations have been carried out, for a molecule with $N=\,100$ base pairs, in a broad temperature range hosting several denaturation steps which start from the AT-rich regions and spread to the GC-base pairs signalling the importance of cooperative effects in twisted geometries. While the melting patterns and the temperature location of the denaturation steps may change for other sequences with different relative weight of GC- and AT- base pairs, our findings regarding the interplay between anharmonicity and twist are independent of the specific sequence.
Melting of heterogeneous DNA generally appears as an overall gradual crossover with presence of closed base pairs portions well inside the denaturation regime and with small entropy jumps. This is largely due to the strong fluctuational effects accounted for by the path integral approach while also heterogeneity smears the transition. The smooth nature of the denaturation is unaltered by the environmental conditions related to the solvent.

\end{document}